\documentclass[11pt,reqno]{article}

\usepackage{amsmath}
\usepackage{amssymb}
\usepackage{amsthm}
\usepackage{bm}
\usepackage[footnotesize,bf]{caption}
\usepackage{color}
\usepackage{eucal}
\usepackage{fullpage}
\usepackage{graphicx}
\usepackage{hhline}
\usepackage{lineno}
\usepackage{makecell}
\usepackage[sc]{mathpazo}
\usepackage{mathrsfs}
\usepackage{mathtools}
\usepackage[numbers,sort&compress,square]{natbib}
\usepackage{setspace}
\usepackage{stmaryrd}
\usepackage[all,cmtip]{xy}

\newcommand*\patchAmsMathEnvironmentForLineno[1]{%
	\expandafter\let\csname old#1\expandafter\endcsname\csname #1\endcsname
	\expandafter\let\csname oldend#1\expandafter\endcsname\csname end#1\endcsname
	\renewenvironment{#1}%
	{\linenomath\csname old#1\endcsname}%
	{\csname oldend#1\endcsname\endlinenomath}}%
\newcommand*\patchBothAmsMathEnvironmentsForLineno[1]{%
	\patchAmsMathEnvironmentForLineno{#1}%
	\patchAmsMathEnvironmentForLineno{#1*}}%
\AtBeginDocument{%
	\patchBothAmsMathEnvironmentsForLineno{equation}%
	\patchBothAmsMathEnvironmentsForLineno{align}%
	\patchBothAmsMathEnvironmentsForLineno{flalign}%
	\patchBothAmsMathEnvironmentsForLineno{alignat}%
	\patchBothAmsMathEnvironmentsForLineno{gather}%
	\patchBothAmsMathEnvironmentsForLineno{multline}%
}

\theoremstyle{definition}

\newtheorem{example}{Example}

\newtheorem{remark}{Remark}
\newtheorem{theorem}{Theorem}

\newcommand{\eq}[1]{\textbf{Eq.~\ref{eq:#1}}}
\newcommand{\eqs}[2]{\textbf{Eqs.~\ref{eq:#1}--\ref{eq:#2}}}
\newcommand{\fig}[1]{\textbf{Fig.~\ref{fig:#1}}}

\newcommand{\thm}[1]{Theorem~\ref{thm:#1}}

\newcommand{\B}{\textrm{G}}
\newcommand{\R}{\textrm{R}}

\title{\begin{center} \bfseries\singlespacing
		The Moran process on 2-chromatic graphs
\end{center}}
\author{\parbox[c]{16cm}{\onehalfspacing \normalsize \centering ~\\[-0.4cm] Kamran Kaveh$^{1}$, Alex McAvoy$^{2}$, Krishnendu Chatterjee$^{3}$, and Martin A. Nowak$^{4,5}$ \\ \quad\\ \footnotesize
		$^{1}$Department of Mathematics, Dartmouth College, Hanover, NH 03755, USA \\
		$^{2}$Department of Mathematics, University of Pennsylvania, Philadelphia, PA~19104 USA \\
		$^{3}$Institute of Science and Technology Austria, Klosterneuburg, Austria \\
		$^{4}$Department of Organismic and Evolutionary Biology, Harvard University, Cambridge, MA 02138, USA \\
		$^{5}$Department of Mathematics, Harvard University, Cambridge, MA 02138, USA\\[0.2cm]}
	\date{}
}

\begin{document}

\allowdisplaybreaks
	
\maketitle

\begin{abstract}
Resources are rarely distributed uniformly within a population. Heterogeneity in the concentration of a drug, the quality of breeding sites, or wealth can all affect evolutionary dynamics. In this study, we represent a collection of properties affecting the fitness at a given location using a color. A green node is rich in resources while a red node is poorer. More colors can represent a broader spectrum of resource qualities. For a population evolving according to the birth-death Moran model, the first question we address is which structures, identified by graph connectivity and graph coloring, are evolutionarily equivalent. We prove that all properly two-colored, undirected, regular graphs are evolutionarily equivalent (where ``properly colored'' means that no two neighbors have the same color). We then compare the effects of background heterogeneity on properly two-colored graphs to those with alternative schemes in which the colors are permuted. Finally, we discuss dynamic coloring as a model for spatiotemporal resource fluctuations, and we illustrate that random dynamic colorings often diminish the effects of background heterogeneity relative to a proper two-coloring.
\end{abstract}


\section*{Introduction}
The survival of a mutant introduced into a population depends on its fitness as well as environmental factors, including the population's spatial structure. Variations in migration patterns or dispersal can be modeled using a graph, where nodes represent individuals and edges represent neighborhoods. On such a graph, a standard measure of a mutant's success is the probability that it takes over the population (``fixes'') \citep{nagylaki1992introduction,durrett2008probability,traulsen2009stochastic,broom2014game,hindersin2015almost,allen2017evolutionary,pavlogiannis:CB:2018,tkadlec:CB:2019,tkadlec:PLOSCB:2020,allen:PLOSCB:2020}. It is known that there is a class of structures, which includes regular graphs, in which all graphs have the same fixation probabilities as those of an unstructured population, a result called the ``isothermal theorem'' \citep{lieberman2005evolutionary}. Our concern here is with an analogue of this result for colored graphs, where each color represents a collection of fitness-influencing resources, in order to work toward a better understand the effects of background fitness heterogeneity on evolutionary dynamics.

Resource heterogeneity naturally results in variations in environmental conditions. For example, the local concentrations of nutrients (sugar) can vary across a population of \textit{E. coli} bacteria. Locations with higher sugar concentration lead to higher reproductive fitness. However, in populations with more than one type of individual, resource heterogeneity can affect the fitness of the different types asymmetrically. Competing \textit{E. coli} strains often metabolize different sugar types (e.g. glucose or lactose). Thus, variations in the concentration of one sugar type predominantly affect the fitness of one \textit{E. coli} strain and not the other \citep{thattai2003metabolic,lambert2014memory}. The evolution of drug resistance is another notable example. The existence of a variable drug distribution across a population (e.g. a drug ``gradient'') can have a strong effect on the onset of drug resistance in microbial evolution \citep{waclaw2015spatial,baym2016spatiotemporal}. Similar observations have been made in the contexts of virus \citep{kepler1998drug,saksena2003reservoirs,singh2010penetration,hermsen2012rapidity,greulich2012mutational,moreno2014imperfect} and cancer \citep{wu2013cell,fu2014spatial} dynamics.

Driven by the ubiquity of heterogeneity within populations, there has been growing interest in understanding how it affects selection in simple mathematical models. Much of this work, ranging from earlier models in population genetics \citep{levene1953genetic,gillespie:AN:1974,frank:AN:1990,gillespie:OUP:1991} to those with more fine-grained spatial structure \citep{masuda2010heterogeneous,hauser2014heterogeneity,maciejewski2014environmental,manem2015modeling,mahdipour2017genotype,farhang2017effect,giaimo2018invasion,farhang:JRSI:2019}, is summarized in a prequel to this study \citep{kaveh2019environmental} (which deals with well-mixed dispersal structures and spatially-modulated fitness). However, a general understanding of the effects of heterogeneous resource distributions within structured populations is still lacking. The subtlety of resource heterogeneity in evolutionary dynamics arises from the interplay of several different parameters: the spatial structure and migration patterns in the population; how each genotype is affected by different concentrations of the resource; the spatial distribution of resources themselves; and finally the fitness of each competing type in the absence of heterogeneity. Due to this complexity, a good deal of the work in this area has been done through numerical simulations of agent-based models with specific structures and fitness distributions.

In this work, we consider heterogeneous resource distributions in graph-structured populations. A distribution of resources within a structured population can be represented by a colored graph, wherein the color of a node represents the resources at that location \citep{maciejewski2014environmental}. Our primary focus is on properly two-colored graphs, which have two distinct node colors together with the property that no two nodes of the same color are neighbors. Each node holds one individual, either a mutant (type $A$) or a resident (type $B$). For evolutionary updating on such a graph, we consider a variant of the well-known Moran process \citep{moran:MPCPS:1958}, which is a birth-death process with one replacement in each time step. A proper two-coloring of the graph in this context implies that an offspring's resources are always different from those of the parent (except, of course, from the trivial case in which the colors have no meaningful effects on the types).

The first question we ask is which population structures and resource distributions result in the same evolutionary dynamics, as measured by the fixation probability of an invading mutant. For properly two-colored graphs in which all nodes of a given color have the same degree (``biregular''), we give an explicit formula for the fixation probability of a rare mutant that is valid for any intensity of selection (\thm{isothermal}). We use this formula to derive a simple condition for when selection favors the mutant type relative to the resident (\thm{selectionCondition}): if $g_{A}$ and $g_{B}$ are the geometric means of the fitness values of $A$ and $B$, respectively, on the two possible colors of the nodes, then selection favors $A$ relative to $B$ if and only if $g_{A}>g_{B}$.

We then consider the effects of resource redistribution within a fixed structure, focusing on the notion of an ``optimal'' distribution of resources with respect to the evolutionary process. On a cycle, a proper two-coloring is typically optimal in the following sense: any reshuffling of colors away from a proper two-coloring leads to a decrease in fixation probability for advantageous mutants and an increase in fixation probability for disadvantageous mutants. We find that this behavior reversed on the star, and it lies somewhere in between on other bipartite graphs with unequal numbers of green and red nodes.

Finally, we explore dynamic graph colorings arising from resource mobility. Resources are redistributed occasionally through a shuffling of the node colors. We find that resource mobility attenuates the effects of background heterogeneity, at least when the level of background heterogeneity is not too large. In some cases, when redistribution occurs at every time step, the effects of background heterogeneity can be completely offset.

\section*{Results}
We begin by modeling the spatial structure of a population with a graph, in which vertices represent locations and edges represent neighbors. The adjacency matrix of this graph, $\left(\Gamma_{ij}\right)_{i,j=1}^{N}$, satisfies $\Gamma_{ij}=1$ if $i$ and $j$ are neighbors and $\Gamma_{ij}=0$ otherwise. There are two types of individuals on the graph, mutants ($A$) and residents ($B$), and each vertex is occupied by exactly one of these two types. When the individual at vertex $i$ reproduces, the offspring is propagated to vertex $j$ with probability $\Gamma_{ij}/\Gamma_{i}$, where $\Gamma_{i}\coloneqq\sum_{k=1}^{N}\Gamma_{ik}$ is the out-degree of vertex $i$. The Moran process \citep{moran:MPCPS:1958} is obtained by choosing one individual to reproduce in each time step, with probability proportional to fitness.

Whereas it is usually assumed that the fitness of $i$ depends on only its type ($A$ or $B$), here we are concerned with fitness that depends on local environmental conditions in addition to an individual's type. To include environmental conditions, we assume that each vertex is assigned a color. For simplicity, we focus on graphs with two possible colors, green and red. For example, a green node might be rich in resources while a red node is poorer. More colors can be used to model a broader spectrum of resource values. We assume the total abundance of resources is constant and does not get degraded over time, which implies that the number of nodes of any given color remains fixed over the course of the evolutionary process.

\begin{figure}
	\centering
	\includegraphics[width=0.6\textwidth]{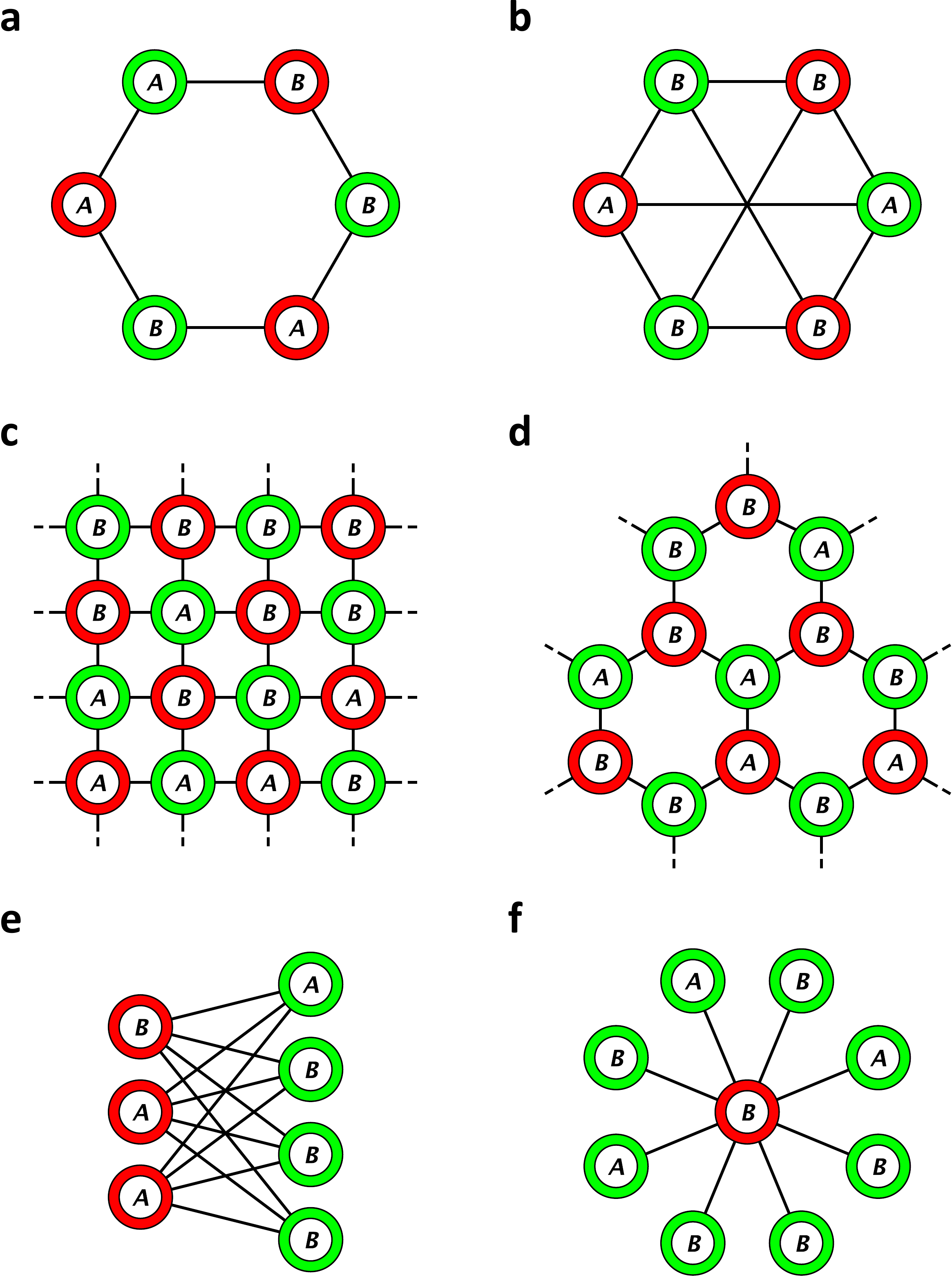}
	\caption{Examples of graphs with proper two-colorings: \textbf{a}, cycle; \textbf{b}, complete, symmetric bipartite graph; \textbf{c}, square lattice (von Neumann neighborhood); \textbf{d}, hexagonal lattice; \textbf{e}, complete, asymmetric bipartite graph; and \textbf{f}, star (an extreme case of \textbf{e}). If the fitness of $A$ and $B$ are $a_{\B}$ and $b_{\B}$ on a green site and $a_{\R}$ and $b_{\R}$ on a red site, respectively, then on each of these graphs we can explicitly calculate the probability that one type replaces the other (\thm{isothermal}).\label{fig:properColorings}}
\end{figure}

A vertex coloring, $\mathcal{C}$, is a function that maps every vertex to a discrete set of values (color set). A coloring map, $\mathcal{C}$, is \emph{proper} if no two nodes of the same color are connected to each other, i.e. $\mathcal{C}\left(i\right)\neq\mathcal{C}\left(j\right)$ whenever $\Gamma_{ij}=1$. A graph is called $n$-colorable if it can be properly colored with $n$ colors. A graph with $N$ vertices is trivially $N$-colorable, but often times a graph can be properly colored with fewer colors. The smallest number of colors for which a proper coloring is possible is known as the \emph{chromatic number} of the graph \citep{bollobas:S:1979}. We are interested in graphs whose chromatic number is $2$, which are also known as ``$2$-chromatic'' or ``bipartite'' graphs (see \fig{properColorings}).

The effects of a color on fitness are described by a \emph{color-to-fitness map}, which maps each color to an ordered set of fitness values, one for each type. For example, in the space of two types, $A$ and $B$, and two colors, green ($\B$) and red ($\R$), the color-to-fitness map can be specified succinctly by a payoff matrix,
\begin{align}
\bordermatrix{
~ & \B & \R \cr
	A & \displaystyle a_{\B} & a_{\R} \cr
	B & \displaystyle b_{\B} & b_{\R}
} .
\end{align}
In other words, on a green node, $A$ has fitness $a_{\B}$ and $B$ has fitness $b_{\B}$. On a red node, $A$ has fitness $a_{\R}$ and $B$ has fitness $b_{\R}$. In evolutionary game theory, such a payoff matrix is usually used to define payoffs for one type against another. Here, this matrix gives the payoffs for a type against the environment.

To quantify the effects of the coloring scheme on evolutionary dynamics, we study the mutant type's fixation probability, $\rho_{A}$, which is the probability that a single, randomly-placed $A$ eventually takes over a background population of type $B$. Similarly, the probability that a single $B$ fixes in a background population of type $A$ is denoted by $\rho_{B}$. Selection is said to favor $A$ relative to $B$ if $\rho_{A}>\rho_{B}$ \citep{proulx:S:2002,king:TPB:2007}. For all graphs of the variety depicted in \fig{properColorings}, we can calculate fixation probabilities explicitly:
\begin{theorem}\label{thm:isothermal}
Consider a graph with a proper two-coloring, with $N_{\B}$ green nodes and $N_{\R}$ red nodes. If every green node has degree $k_{\B}$ and every red node has degree $k_{\R}$, then the mean fixation probability of $A$ appearing uniformly at random in a background population of $B$ is
\begin{align}
\rho_{A} &= \frac{\displaystyle 1-\frac{1}{N_{\B}+N_{\R}}\left\{N_{\B}\frac{b_{\R}\left(a_{\R}k_{\B}+b_{\B}k_{\R}\right)}{a_{\R}\left(a_{\B}k_{\R}+b_{\R}k_{\B}\right)}+N_{\R}\frac{b_{\B}\left(a_{\B}k_{\R}+b_{\R}k_{\B}\right)}{a_{\B}\left(a_{\R}k_{\B}+b_{\B}k_{\R}\right)}\right\}}{\displaystyle 1-\left(\frac{b_{\R}\left(a_{\R}k_{\B}+b_{\B}k_{\R}\right)}{a_{\R}\left(a_{\B}k_{\R}+b_{\R}k_{\B}\right)}\right)^{N_{\B}}\left(\frac{b_{\B}\left(a_{\B}k_{\R}+b_{\R}k_{\B}\right)}{a_{\B}\left(a_{\R}k_{\B}+b_{\B}k_{\R}\right)}\right)^{N_{\R}}} . \label{eq:fp_bipartite}
\end{align}
Similarly, the mean fixation probability of $B$ appearing uniformly at random in a background population of $A$ is
\begin{align}
\rho_{B} &= \frac{\displaystyle 1-\frac{1}{N_{\B}+N_{\R}}\left\{N_{\B}\frac{a_{\R}\left(b_{\R}k_{\B}+a_{\B}k_{\R}\right)}{b_{\R}\left(b_{\B}k_{\R}+a_{\R}k_{\B}\right)}+N_{\R}\frac{a_{\B}\left(b_{\B}k_{\R}+a_{\R}k_{\B}\right)}{b_{\B}\left(a_{\R}k_{\B}+a_{\B}k_{\R}\right)}\right\}}{\displaystyle 1-\left(\frac{a_{\R}\left(b_{\R}k_{\B}+a_{\B}k_{\R}\right)}{b_{\R}\left(b_{\B}k_{\R}+a_{\R}k_{\B}\right)}\right)^{N_{\B}}\left(\frac{a_{\B}\left(b_{\B}k_{\R}+a_{\R}k_{\B}\right)}{b_{\B}\left(b_{\R}k_{\B}+a_{\B}k_{\R}\right)}\right)^{N_{\R}}} .
\end{align}
\end{theorem}

\begin{remark}
Under the assumptions of \thm{isothermal}, we must have $N_{\B}k_{\B}=N_{\R}k_{\R}$. Thus, $N_{\B}=N_{\R}$ if and only if $k_{\B}=k_{\R}$ (the graph is regular), in which case $\rho_{A}$ is independent of the degree.
\end{remark}

From \thm{isothermal}, we can derive a simple condition for selection to favor $A$:
\begin{theorem}\label{thm:selectionCondition}
For a graph satisfying the hypotheses of \thm{isothermal},
\begin{align}
\rho_{A} > 1/N > \rho_{B} &\iff \rho_{A} > \rho_{B} \iff a_{\B}a_{\R} > b_{\B}b_{\R} .
\end{align}
\end{theorem}

We give proofs of Theorems~\ref{thm:isothermal}--\ref{thm:selectionCondition}, as well as an extension to weighted, directed graphs, in the Appendix.

\subsection*{Parametrizing heterogeneity}
\begin{figure}
	\centering
	\includegraphics[width=0.8\textwidth]{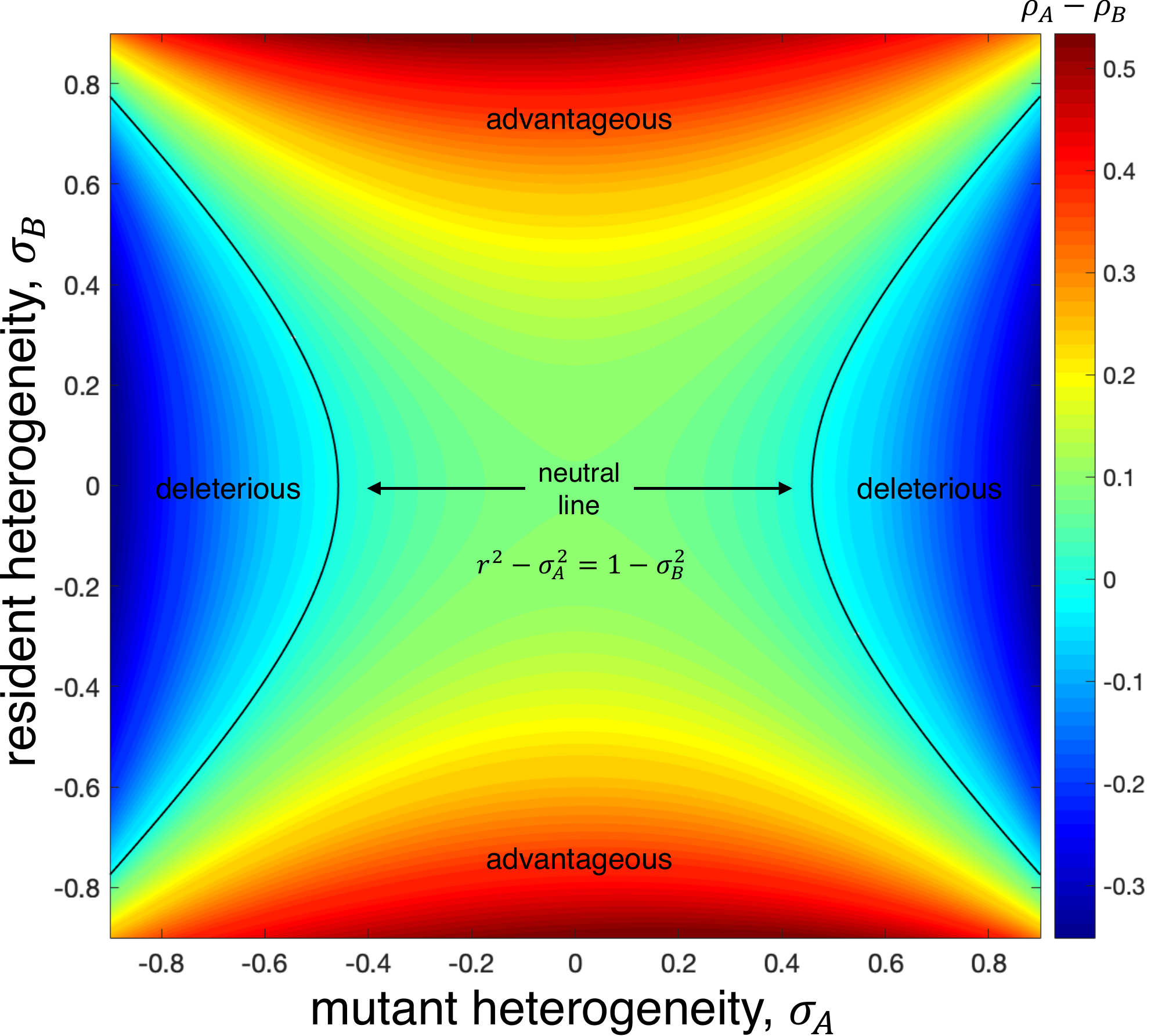}
	\caption{Selection condition when $a_{\B}=r+\sigma_{A}$, $a_{\R}=r-\sigma_{A}$, $b_{\B}=1+\sigma_{B}$, and $b_{\R}=1-\sigma_{B}$. The graph is bipartite, regular, and properly two-colored with $N_{\B}=N_{\R}=50$. The difference between fixation probabilities of the two types, $\rho_{A}-\rho_{B}$, is shown for several values of $\sigma_{A}$ and $\sigma_{B}$ when $r=1.5$. Warmer colors represent greater differences between the fixation probabilities, $\rho_{A}-\rho_{B}$. $A$ is neutral relative to $B$ if and only if $r^{2}-\sigma_{A}^{2}=1-\sigma_{B}^{2}$.\label{fig:selectionCondition}}
\end{figure}

Let $r_{A}$ and $r_{B}$ represent the mean fitness values of $A$ and $B$, respectively, over all locations in the population, i.e.
\begin{subequations}\label{eq:meanFitness}
\begin{align}
r_{A} &= \frac{N_{\B}a_{\B}+N_{\R}a_{\R}}{N_{\B}+N_{\R}} ; \\
r_{B} &= \frac{N_{\B}b_{\B}+N_{\R}b_{\R}}{N_{\B}+N_{\R}} .
\end{align}
\end{subequations}
The difference between the fitness of $A$ on a green node, $a_{\B}$, and the mean fitness of $A$, $r_{A}$, is denoted $\sigma_{A}^{\B}\coloneqq a_{\B}-r_{A}$. Similarly, for type $B$ on a green node, we let $\sigma_{B}^{\B}\coloneqq b_{\B}-r_{B}$. Since we usually consider green nodes to be beneficial, it is typically the case that $\sigma_{A}^{\B},\sigma_{B}^{\B}\geqslant 0$. In this way, these two parameters quantify the (beneficial) effects of green nodes on the two types. On the other hand, red nodes are usually deleterious, which means that the quantities $\sigma_{A}^{\R}\coloneqq r_{A}-a_{\R}$ and $\sigma_{B}^{\R}\coloneqq r_{B}-b_{\R}$ are at least zero and quantify the degree to which red nodes are harmful to the two types. By \eq{meanFitness}, we must have $N_{\B}\sigma_{A}^{\B}=N_{\R}\sigma_{A}^{\R}$ and $N_{\B}\sigma_{B}^{\B}=N_{\R}\sigma_{B}^{\R}$. In particular, the heterogeneity can be parametrized by just two values, $\sigma_{A}\coloneqq\sigma_{A}^{\B}$ (mutant heterogeneity) and $\sigma_{B}\coloneqq\sigma_{B}^{\B}$ (resident heterogeneity). (Note, however, that these parameters are added to fitness on green nodes and subtracted from fitness on red nodes.) By \thm{selectionCondition}, the selection condition is
\begin{align}
\rho_{A} > 1/N > \rho_{B} &\iff \rho_{A} > \rho_{B} \iff \left(r_{A}+\sigma_{A}\right)\left(N_{\R}r_{A}-N_{\B}\sigma_{A}\right) > \left(r_{B}+\sigma_{B}\right)\left(N_{\R}r_{B}-N_{\B}\sigma_{B}\right) . \label{eq:simplifiedCondition}
\end{align}
\fig{selectionCondition} illustrates this selection condition when $N_{\B}=N_{\R}=50$, $r_{A}=1.5$, and $r_{B}=1$.

Several natural special cases emerge from this general accounting of heterogeneity:
\begin{itemize}
	
\item[\emph{(i)}] symmetric environmental interactions: $\sigma_{A}=\sigma_{B}$;

\item[\emph{(ii)}] asymmetric environmental interactions: $\sigma_{A}=-\sigma_{B}$;

\item[\emph{(iii)}] mutant heterogeneity: $\sigma_{B}=0$;

\item[\emph{(iv)}] resident heterogeneity: $\sigma_{A}=0$.

\end{itemize}
Note that each of these cases is a one-parameter model of heterogeneity. Our main focus here is on \emph{(i)}, where the coloring affects the fitness of the two types symmetrically. We refer to this case as ``heterogeneous background fitness'' and use $\sigma$ to denote the parameter $\sigma_{A}$. For simplicity, we also let $r_{B}=1$ and $r\coloneqq r_{A}$.

\begin{figure}
	\centering
	\includegraphics[width=0.8\textwidth]{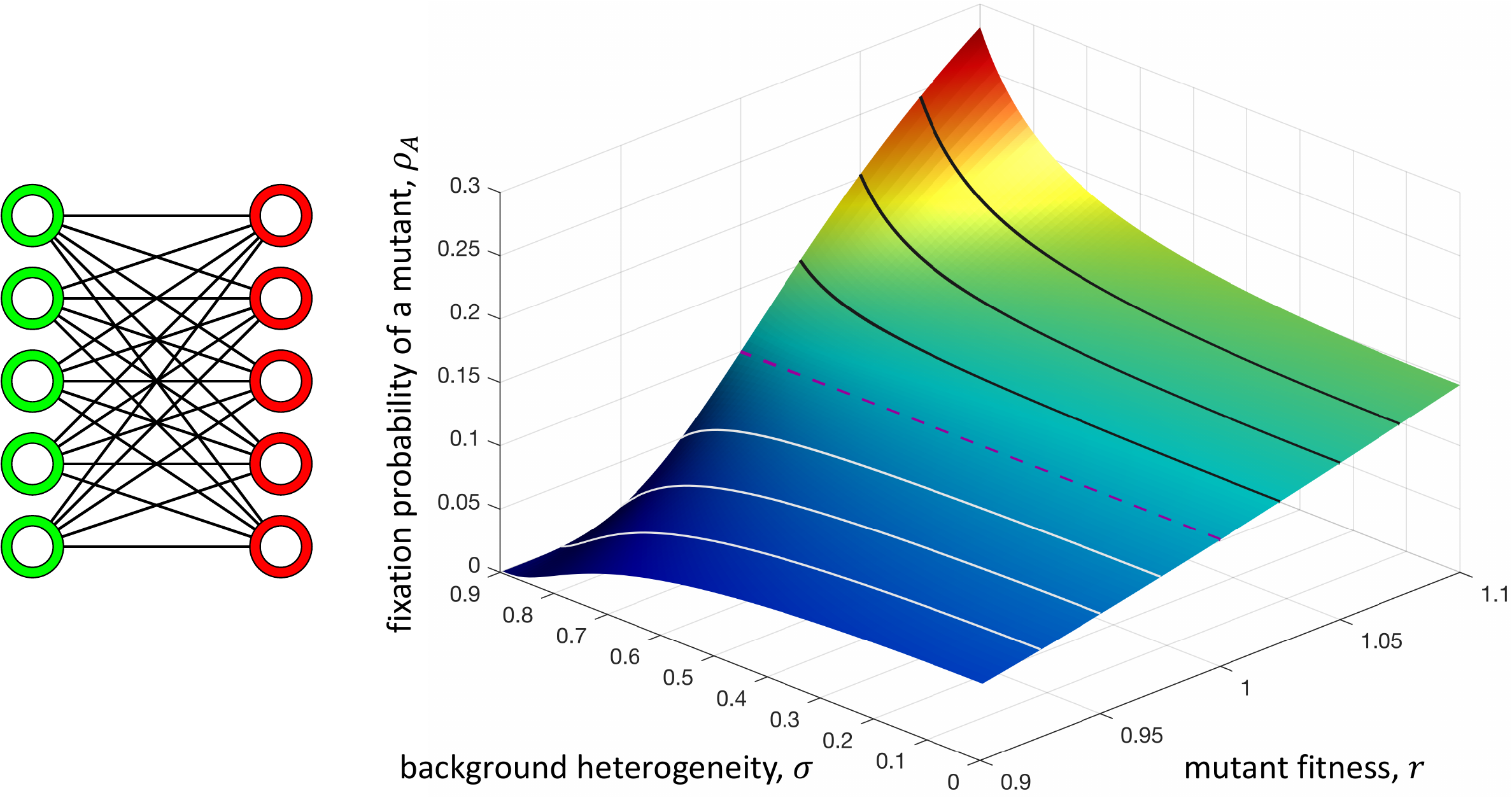}
	\caption{Effects of background heterogeneity on a complete bipartite graph with $N_{\B}=N_{\R}=5$. Background heterogeneity increases the fixation probability of an advantageous mutant (black lines) and decreases that of a disadvantageous mutant (white lines). These effects are monotonic in background heterogeneity, $\sigma$. By \thm{isothermal}, this behavior is identical to that of a regular (not necessarily complete) properly two-colored graph with equal numbers of green and red nodes.\label{fig:amplifierByColor}}
\end{figure}

For example, if $N_{\B}=N_{\R}=N/2$, then, under heterogeneous background fitness, \eq{fp_bipartite} simplifies to
\begin{align}
\rho_{A} &= \frac{\displaystyle r\left(r-1\right)}{\displaystyle \left(r^{2}-\sigma^{2}\right)\left(1-\left(\frac{1-\sigma^{2}}{r^{2}-\sigma^{2}}\right)^{N/2}\right)} .
\end{align}
From this expression, we see that the fixation probability of a mutant is an increasing function of $\sigma$ when the mutant is advantageous ($r>1$), a decreasing function of $\sigma$ when the mutant is disadvantageous ($r<1$), and independent of $\sigma$ when the mutant is neutral ($r=1$). This result is depicted in \fig{amplifierByColor} on a graph with $N_{\B}=N_{\R}=5$.

\subsection*{Resource redistribution and optimal colorings}
On a colored graph, shuffling the node colors does not change the overall number of each color present in the population. In other words, shuffling leaves the total resource value constant. Having established formulas for fixation probabilities on biregular, properly two-colored graphs, we now turn to fixation probabilities on graphs with various permutations of proper two-colorings. This model of permuted colorings is related to (and inspired by) the work of \citet{mahdipour2017genotype}, which explores how increasing the standard deviation of a bimodel fitness distribution affects fixation probabilities in simple structured populations. In the model we consider, we permute the colors at the beginning of the process and then leave them fixed for the remainder.

\begin{figure}
	\centering
	\includegraphics[width=1.0\textwidth]{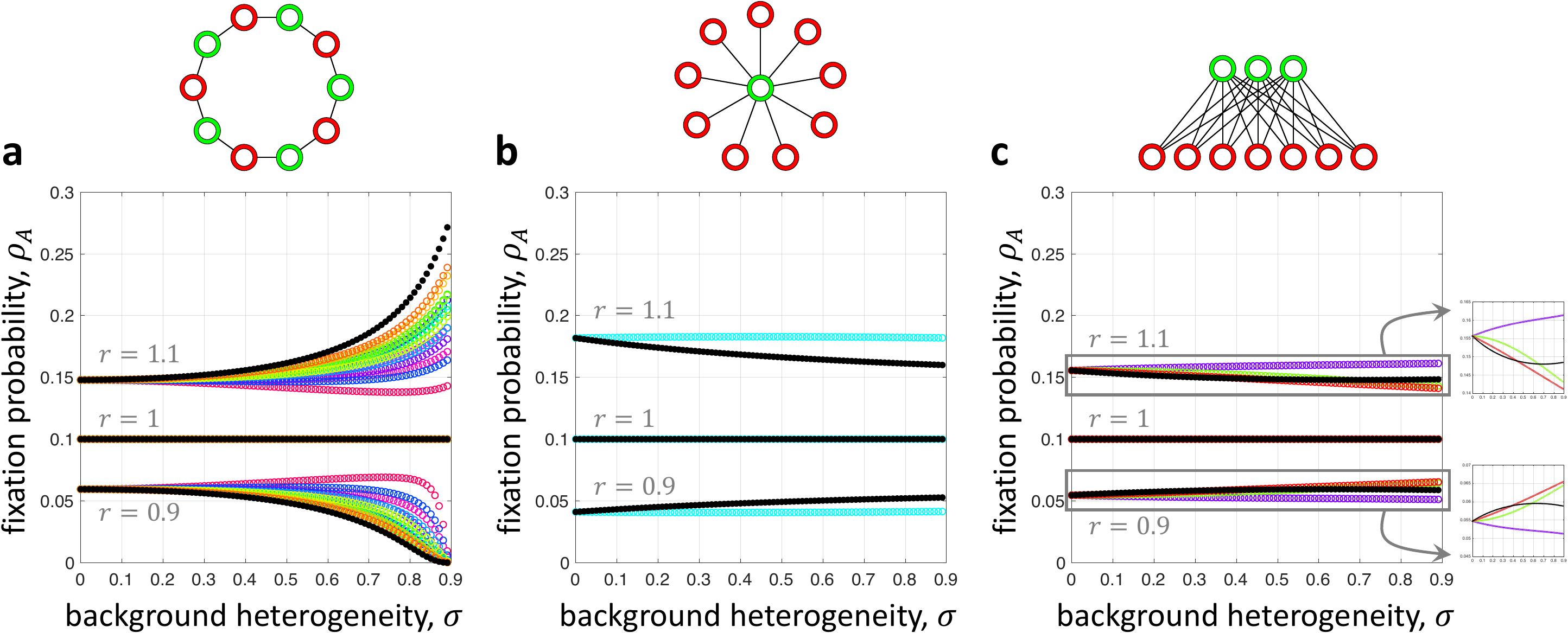}
	\caption{Effects of background heterogeneity for alternative colorings on two-colorable graphs. In each panel, we plot fixation probability against background heterogeneity for all (non-isomorphic) permutations of the proper two-coloring shown at the top. The proper two-coloring in each case is depicted in black, which is given by \eq{fp_bipartite}. \textbf{a}, On the cycle, the proper two-coloring is ``optimal'' in the sense that it gives the maximum fixation probability for an advantageous mutant and the minimum fixation probability for a disadvantageous mutant. \textbf{b}, On the star, we observe the opposite behavior, with the proper two-coloring giving the minimum fixation probability for an advantageous mutant and the maximum for a disadvantageous mutant. \textbf{c}, On a complete bipartite graph with $N_{\B}\neq N_{\R}$ (shown here with $N_{\B}=3$ and $N_{\R}=7$), a mixture of these two results is possible. In particular, there need not be a coloring that is ``optimal'' for all levels of background heterogeneity. The fixation probabilities in all panels were approximated by building transition matrices for each process and looking at the exact distribution after $10^{7}$ steps.\label{fig:redistribution}}
\end{figure}

On the cycle, we find that shuffling the colors away from a proper two-coloring attenuates the effects of background heterogeneity (\fig{redistribution}a). On the star with $N_{\B}=1$ and $N_{\R}=N-1$, there are just two non-isomorphic colorings, one proper and one non-proper. Moving from the proper coloring to the non-proper coloring increases an advantageous mutant's fixation probability and decreases that of a disadvantageous mutant (\fig{redistribution}b), which is strictly the opposite of the behavior observed for the cycle. In between these two population structures is a complete bipartite graph with $1<N_{\B}<N_{\R}<N$. On such a structure, the effects of moving away from a proper coloring are not quite as uniform: for some values of background heterogeneity, an advantageous mutant's fixation probability is increased, while it is decreased for other values of $\sigma$ (\fig{redistribution}c).

\subsection*{Dynamic coloring and resource mobility}
So far, we have considered the case in which resources (and thus environmental conditions) are spatially distributed across the population but fixed over time. In many cases, resources and local conditions change over time as well. For example, an area with rich soil or a good climate can deteriorate with time, leading to poorer conditions (and vice versa). Resource agents, such as nutrients in a heterogeneous population of bacteria, can be mobile and diffuse across an evolving population. It is therefore natural to consider the effects of heterogeneity on evolutionary dynamics when the distribution of resources is itself dynamic.

Spatial and temporal heterogeneity have been observed (separately) to have quite different effects on the fixation probability of a mutant \citep{farhang:JRSI:2019}. Here, we consider a model that combines these two kinds of heterogeneity. For simplicity, we focus on a graph-structured population with two colors. We consider a simple model of dynamic coloring in which, at each time step, the colors on the graph are shuffled according to a random permutation. To control the speed of resource movements, we assume that a shuffling happens with some fixed probability, $p$. With probability $1-p$, there is no movement and the color scheme remains the same as it was in the previous time step. This model is similar to others involving ``motion'' \citep{krieger:JRSI:2017,herrerias:SR:2018}, except here it acts on resources rather than traits.

\begin{figure}
	\centering
	\includegraphics[width=1.0\textwidth]{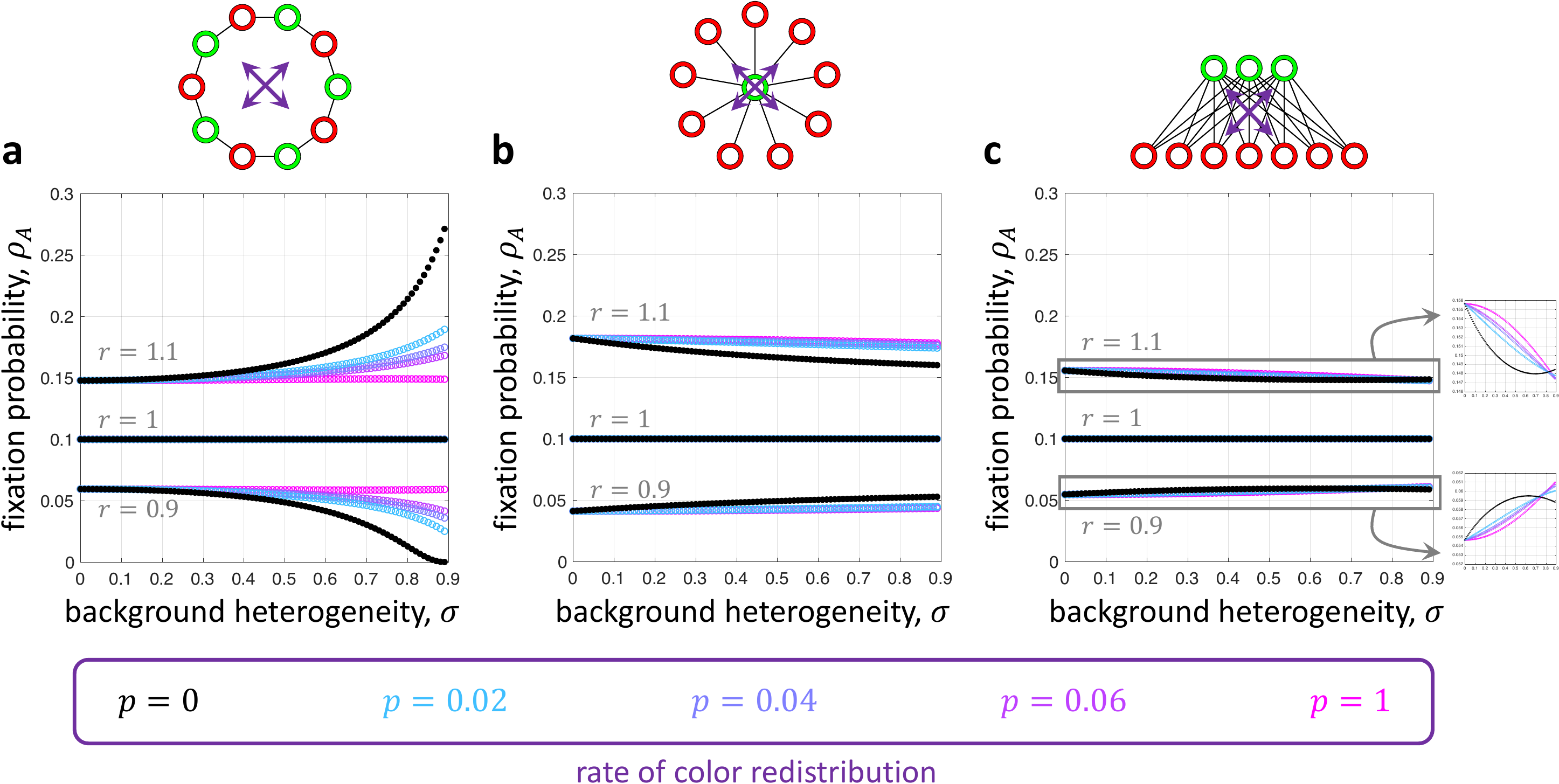}
	\caption{Fixation probability in the presence of dynamic resources on bipartite graphs. $\rho_{A}$ is shown here as a function of resource heterogeneity, $\sigma$, for several values of the resource redistribution rate, $p$, and mutant mean fitness, $r$. The population initially has a proper two-coloring. At each time step, with probability $p$ the colors are shuffled according to permutation chosen uniformly at random. With probability $1-p$, the coloring is not changed in that time step. In \textbf{a} and \textbf{b}, high environmental fluctuations (i.e. large $p$) attenuate the effects of background heterogeneity relative to the initial distribution of resources at $p=0$ (corresponding to a proper two-coloring, shown in black and given by \eq{fp_bipartite}). In \textbf{c}, this behavior holds for all but the highest levels of background heterogeneity. The fixation probabilities in all panels were approximated by building transition matrices for each process and looking at the exact distribution after $10^{7}$ steps.\label{fig:dynamicColoring}}
\end{figure}

As examples, we consider an undirected cycle, star, and another complete bipartite graph with $N_{\B}\neq N_{\R}$ for several values of the shuffling rate, $p$. \fig{dynamicColoring} shows the fixation probability as a function of $\sigma$ starting from a two-colored assignment. For $p=0$, \thm{isothermal} gives the exact fixation probability of $A$. As $p$ increases, the effects of resource heterogeneity weaken, as long as the level of background heterogeneity is not too high. However, when background heterogeneity is sufficiently strong and the population is asymmetric, resource mobility can actually strengthen the effects of heterogeneity on a mutant's fixation probability, relative to the fixed proper two-coloring when $p=0$ (\fig{dynamicColoring}c).

\section*{Discussion}
Sources of heterogeneity abound in nature, at all scales. The distribution of a drug across a tumor can be highly heterogeneous due to non-uniformity in the vasculature as well as in the tumor's tissue-like structure \citep{fu2014spatial}. A good breeding site may give a bird an advantage, which is sometimes connected to its own behavior \citep{misenhelter2000choices}. A good school district can have a profound impact on one's education and career progression \citep{cullen2005impact}, and inherited wealth may positively affect reproductive success \citep{essock1984reproductive}. On the flip side, many modeling scenarios require environmental heterogeneity together with frequency-dependent fitness (i.e. ``games''). Provided selection is sufficiently weak, analytical results on the effects of heterogeneity on an evolving population can be derived in that setting as well \citep[see][]{mcavoy:NHB:2020}. In contrast, the analytical results presented here hold for \textit{any} selection strength, with the trade-off being that they require the assumptions of frequency-independent fitness and sufficient regularity in the population structure and coloring scheme.

While these assumptions are somewhat restrictive, they allow for a formal analysis of the model that reveals several interesting effects of background heterogeneity. When heterogeneity affects the two types ($A$ and $B$) in the same way, it can increase an advantageous mutant's fixation probability and decrease that of a disadvantageous individual. As the graph becomes more heterogeneous, here in the form of having unequal numbers of green and red nodes, both the effects of heterogeneity and the relationship between proper and non-proper colorings on the graph grow more nuanced. Our focus here was on an analogue of the isothermal theorem \citep{lieberman2005evolutionary} for colored graphs, but other classes of colorings on more complicated heterogeneous graphs would also be relevant for understanding the evolutionary implications of background heterogeneity.

In addition to earlier work in population genetics, several steps have recently been taken to better understand the roles of heterogeneity in evolutionary dynamics \citep{manem2014spatial,manem2015modeling,hauser2014heterogeneity,farhang2017effect,mahdipour2017genotype,kaveh2019environmental,farhang:JRSI:2019}. However, a complete picture remains elusive. Due to the complexity of analyzing models that account for heterogeneity, such an understanding likely will not emerge for some time. Given the rapid development of the Moran model in structured populations over the past decade-and-a-half, we see generalizations involving both spatial and temporal resource heterogeneity to be natural next steps, both for their mathematical intrigue and their applicability to the heterogeneity that is present in nearly every natural population.

\section*{Methods}

\subsection*{Calculating fixation probabilities}
Consider a colored graph of size $N$ in which each node is occupied by a single individual of type $A$ (mutant) or $B$ (resident). The fitness of a type depends on both the type itself and the color of its node. We denote the fitness of $A$ and $B$ at location $i$ by $a_{i}$ and $b_{i}$, respectively. The fitness values belong to a finite set defined by the number of colors. For example in case of two colors, green and red, the possible fitness sets for mutants are two values, $a_{i}\in\left\{a_{\B},a_{\R}\right\}$, where $a_{\B}$ and $a_{\R}$ denote the fitness of $A$ on green and red sites, respectively. Our focus is on graphs with two colors, having $N_{\B}$ green nodes and $N_{\R}$ red nodes.

\subsubsection*{Directed, weighted graphs}
The probability that an offspring produced at site $i$ replaces an individual residing at a neighboring location, $j$, is determined by a directed, weighted dispersal graph with matrix $\left(w_{ij}\right)_{i,j=1}^{N}$, having no self-loops (meaning $w_{ii}=0$ for $i=1,\dots ,N$). In a standard extension of the Moran model to graphs, an individual at location $i$ replaces $j$ with probability proportional to the product of $i$'s fitness and the edge weight $w_{ij}$ \citep{lieberman2005evolutionary}. The overall structure of the population is defined by both this dispersal graph and the coloring of the nodes.

A state of the process can be described by a vector, $\mathbf{x}\in\left\{0,1\right\}^{N}$, where $x_{i}=1$ (resp. $x_{i}=0$) indicates that the individual at location $i$ has type $A$ (resp. $B$). The total fitness of type $B$ in the population is $\sum_{j=1}^{N}\left(1-x_{j}\right) b_{j}$. Multiplying each location $j$ by the edge weight $w_{ji}$ if $i$ has type $A$ (and by $0$ otherwise), we see that the probability of losing a mutant in location $i$ is proportional to $\sum_{j=1}^{N}\Big(\left(1-x_{j}\right) b_{j}\Big)\Big(w_{ji}x_{i}\Big)$, where the term $\left(1-x_{j}\right) b_{j}$ corresponds to the birth of type $B$ and $w_{ji}x_{i}$ corresponds to the death of a neighboring type $A$. Similarly, the probability of gaining a mutant at location $i$ is proportional to $\sum_{j=1}^{N}\Big(x_{j}a_{j}\Big)\Big(w_{ji}\left(1-x_{i}\right)\Big)$. Finally, the probability that location $i$ neither gains nor loses a mutant is proportional to
\begin{align}
\sum_{j=1}^{N}\left( x_{i}x_{j}a_{j} + \left(1-x_{i}\right)\left(1-x_{j}\right) b_{j} \right) w_{ji} .
\end{align}
It follows that the probabilities of losing and gaining a mutant in state $\mathbf{x}$ are
\begin{subequations}\label{eq:transition}
\begin{align}
P_{i}^{-}\left(\mathbf{x}\right) &= \frac{\sum_{j=1}^{N}x_{i}\left(1-x_{j}\right) b_{j}w_{ji}}{\sum_{j,k=1}^{N}\left(x_{j}a_{j}+\left(1-x_{j}\right) b_{j}\right) w_{jk}} ; \\
P_{i}^{+}\left(\mathbf{x}\right) &= \frac{\sum_{j=1}^{N}\left(1-x_{i}\right) x_{j}a_{j}w_{ji}}{\sum_{j,k=1}^{N}\left(x_{j}a_{j}+\left(1-x_{j}\right) b_{j}\right) w_{jk}} ,
\end{align}
\end{subequations}
respectively. Conditioned on a change in mutant frequency (meaning the process cannot remain in the same state, provided this state is non-absorbing), the transition probabilities are given by
\begin{subequations}\label{eq:Qdefinition}
\begin{align}
Q_{i}^{-}\left(\mathbf{x}\right) &= \frac{P_{i}^{-}\left(\mathbf{x}\right)}{\sum_{k=1}^{N}\left(P_{k}^{-}\left(\mathbf{x}\right) +P_{k}^{+}\left(\mathbf{x}\right)\right)} = \frac{\sum_{j=1}^{N}x_{i}\left(1-x_{j}\right) b_{j}w_{ji}}{\sum_{j,k=1}^{N}\left(\left(1-x_{k}\right) x_{j}a_{j}w_{jk}+x_{k}\left(1-x_{j}\right) b_{j}w_{jk}\right)} ; \\
Q_{i}^{+}\left(\mathbf{x}\right) &= \frac{P_{i}^{+}\left(\mathbf{x}\right)}{\sum_{k=1}^{N}\left(P_{k}^{-}\left(\mathbf{x}\right) +P_{k}^{+}\left(\mathbf{x}\right)\right)} = \frac{\sum_{j=1}^{N}\left(1-x_{i}\right) x_{j}a_{j}w_{ji}}{\sum_{j,k=1}^{N}\left(\left(1-x_{k}\right) x_{j}a_{j}w_{jk}+x_{k}\left(1-x_{j}\right) b_{j}w_{jk}\right)} .
\end{align}
\end{subequations}

For $\zeta_{1},\dots ,\zeta_{N}\in\mathbb{R}$ and $\mathbf{x}\in\left\{0,1\right\}^{N}$, let $\bm{\zeta}^{\mathbf{x}}$ denote the product $\zeta_{1}^{x_{1}}\cdots\zeta_{N}^{x_{N}}$. Let $\left\{X_{n}\right\}_{n\geqslant 0}$ represent the state of the process, with $X_{n}=\mathbf{x}$ indicating that the state of the population at time $n$ is $\mathbf{x}$. Due to the well-known connection between fixation probabilities and martingales \citep[see][]{monk2014martingales}, we are first interested in finding $\zeta_{1},\dots ,\zeta_{N}\in\mathbb{R}$ for which $\left\{\bm{\zeta}^{X_{n}}\right\}_{n\geqslant 0}$ is a martingale with respect to $\left\{X_{n}\right\}_{n\geqslant 0}$, which means that $\mathbb{E}\left[\bm{\zeta}^{X_{n+1}}\ \mid\ X_{0},X_{1},\dots ,X_{n}\right] = \bm{\zeta}^{X_{n}}$ for every $n\geqslant 0$. Since $\left\{X_{n}\right\}_{n\geqslant 0}$ is a Markov chain and the law of $X_{n+1}$ depends on only $X_{n}$, it suffices to find $\zeta_{1},\dots ,\zeta_{N}\in\mathbb{R}$ such that $\mathbb{E}\left[\bm{\zeta}^{X_{n+1}}\ \mid\ X_{n}\right] = \bm{\zeta}^{X_{n}}$. Using \eq{transition}, we see that
\begin{align}
\mathbb{E}\left[\bm{\zeta}^{X_{n+1}}\ \mid\ X_{n}=\mathbf{x}\right] &= \sum_{i=1}^{N} \left( P_{i}^{-}\left(\mathbf{x}\right)\bm{\zeta}^{\mathbf{x}}\zeta_{i}^{-1} + P_{i}^{+}\left(\mathbf{x}\right)\bm{\zeta}^{\mathbf{x}}\zeta_{i} \right) + \left(1-\sum_{i=1}^{N} \left( P_{i}^{-}\left(\mathbf{x}\right) + P_{i}^{+}\left(\mathbf{x}\right) \right)\right)\bm{\zeta}^{\mathbf{x}} .
\end{align}
Therefore, by \eqs{transition}{Qdefinition}, the equation $\mathbb{E}\left[\bm{\zeta}^{X_{n+1}}\ \mid\ X_{n}=\mathbf{x}\right] = \bm{\zeta}^{\mathbf{x}}$ is equivalent to
\begin{align}
\sum_{i=1}^{N} &\left( P_{i}^{-}\left(\mathbf{x}\right)\bm{\zeta}^{\mathbf{x}}\zeta_{i}^{-1} + P_{i}^{+}\left(\mathbf{x}\right)\bm{\zeta}^{\mathbf{x}}\zeta_{i} \right) + \left(1-\sum_{i=1}^{N} \left( P_{i}^{-}\left(\mathbf{x}\right) + P_{i}^{+}\left(\mathbf{x}\right) \right)\right)\bm{\zeta}^{\mathbf{x}} = \bm{\zeta}^{\mathbf{x}} \nonumber \\
&\iff \sum_{i=1}^{N} \left( Q_{i}^{-}\left(\mathbf{x}\right)\zeta_{i}^{-1} + Q_{i}^{+}\left(\mathbf{x}\right)\zeta_{i} \right) = 1 . \label{eq:Qmartingale}
\end{align}
By \eq{Qdefinition}, we see that \eq{Qmartingale} holds for every $\mathbf{x}\in\left\{0,1\right\}^{N}$ if and only if
\begin{align}
0 &= \sum_{i,j=1}^{N}x_{i}\left(1-x_{j}\right) b_{j}w_{ji}\left(\zeta_{i}^{-1}-1\right) +\sum_{i,j=1}^{N}\left(1-x_{i}\right) x_{j}a_{j}w_{ji}\left(\zeta_{i}-1\right) \nonumber \\
&= \sum_{i=1}^{N} \left[ \sum_{j=1}^{N} b_{j}w_{ji}\left(\zeta_{i}^{-1}-1\right) + \sum_{j=1}^{N} a_{i}w_{ij}\left(\zeta_{j}-1\right) \right] x_{i} \nonumber \\
&\quad - \sum_{i<j} \left[ b_{j}w_{ji}\left(\zeta_{i}^{-1}-1\right) + a_{j}w_{ji}\left(\zeta_{i}-1\right) + b_{i}w_{ij}\left(\zeta_{j}^{-1}-1\right) + a_{i}w_{ij}\left(\zeta_{j}-1\right) \right] x_{i}x_{j} \label{eq:xMartingale}
\end{align}
for every $\mathbf{x}\in\left\{0,1\right\}^{N}$. The right-hand side of \eq{xMartingale} is a multi-linear polynomial representation of a pseudo-Boolean function (i.e. a mapping $\left\{0,1\right\}^{N}\rightarrow\mathbb{R}$) that is identically zero, so its coefficients must also all be zero by uniqueness of this representation \citep{hammer:S:1968,boros:DAM:2002}. Therefore, for every $i,j=1,\dots ,N$, we obtain the system of equations
\begin{subequations}\label{eq:martingaleLinearQuadratic}
\begin{align}
\sum_{j=1}^{N}b_{j}w_{ji}\left(\zeta_{i}^{-1}-1\right) +\sum_{j=1}^{N}a_{i}w_{ij}\left(\zeta_{j}-1\right) &= 0 ; \quad\left(\textrm{linear terms}\right) \label{eq:martingaleLinear} \\
b_{j}w_{ji}\left(\zeta_{i}^{-1}-1\right) +a_{j}w_{ji}\left(\zeta_{i}-1\right) +b_{i}w_{ij}\left(\zeta_{j}^{-1}-1\right) +a_{i}w_{ij}\left(\zeta_{j}-1\right) &= 0 . \quad\left(\textrm{quadratic terms}\right) \label{eq:martingaleQuadratic}
\end{align}
\end{subequations}

Let $\rho_{A}\left(\mathbf{x}\right)$ denote the fixation probability of type $A$ when starting in the initial configuration $\mathbf{x}\in\left\{0,1\right\}^{N}$. If there exist $\zeta_{1},\dots ,\zeta_{N}\in\mathbb{R}$ solving \eq{martingaleLinearQuadratic}, then it is well-known that the martingale property yields
\begin{align}
\zeta_{1}^{x_{1}}\cdots\zeta_{N}^{x_{N}} &= \left(1-\rho_{A}\left(\mathbf{x}\right)\right) + \rho_{A}\left(\mathbf{x}\right) \zeta_{1}\cdots\zeta_{N}
\end{align}
\citep[see][]{monk2014martingales,kaveh2019environmental}. Solving for $\rho_{A}\left(\mathbf{x}\right)$ gives a fixation probability of
\begin{align}
\rho_{A}\left(\mathbf{x}\right) &= \frac{1-\zeta_{1}^{x_{1}}\cdots\zeta_{N}^{x_{N}}}{1-\zeta_{1}\cdots\zeta_{N}} . \label{eq:fp}
\end{align}
Thus, in what follows, we can reduce the problem of calculating fixation probabilities to finding $\zeta_{1},\dots ,\zeta_{N}\in\mathbb{R}$ that satisfy \eq{martingaleLinearQuadratic}.

In general, there need not be a solution to \eq{martingaleLinearQuadratic}. However, with a few simplifying assumptions, we can reduce this system of equations to something more manageable. Specifically, we seek a solution to this system for which $\zeta_{i}$ depends on only the color of $i$. For every node $i$, we use the notation $i\sim\B$ (resp. $i\sim\R$) to denote that $i$ is colored green (resp. red). The first assumption is that the graph is properly two-colored, which means that $w_{ij}=0$ whenever $i,j\sim\B$ or $i,j\sim\R$. The second assumption is that the indegree is the same for all nodes of the same color, as is the outdegree. Let $w_{\B}^{\textrm{in}}$ and $w_{\B}^{\textrm{out}}$ (resp. $w_{\R}^{\textrm{in}}$ and $w_{\R}^{\textrm{out}}$) denote the indegree and outdegree, respectively, of green (resp. red) nodes. Under these assumptions, such a solution to \eq{martingaleLinear} requires
\begin{subequations}
\begin{align}
b_{\R}\left(\zeta_{\B}^{-1}-1\right) w_{\B}^{\textrm{in}} +a_{\B} \left(\zeta_{\R}-1\right) w_{\B}^{\textrm{out}} &= 0 ; \\
b_{\B}\left(\zeta_{\R}^{-1}-1\right) w_{\R}^{\textrm{in}} +a_{\R} \left(\zeta_{\B}-1\right) w_{\R}^{\textrm{out}} &= 0 .
\end{align}
\end{subequations}
It is straightforward to see that the unique solution to these equations is given by
\begin{subequations}\label{eq:zetaBzetaR}
\begin{align}
\zeta_{\B} &= \frac{b_{\R}w_{\B}^{\textrm{in}}\left(a_{\R}w_{\R}^{\textrm{out}}+b_{\B}w_{\R}^{\textrm{in}}\right)}{a_{\R}w_{\R}^{\textrm{out}}\left(a_{\B}w_{\B}^{\textrm{out}}+b_{\R}w_{\B}^{\textrm{in}}\right)} ; \\
\zeta_{\R} &= \frac{b_{\B}w_{\R}^{\textrm{in}}\left(a_{\B}w_{\B}^{\textrm{out}}+b_{\R}w_{\B}^{\textrm{in}}\right)}{a_{\B}w_{\B}^{\textrm{out}}\left(a_{\R}w_{\R}^{\textrm{out}}+b_{\B}w_{\R}^{\textrm{in}}\right)} .
\end{align}
\end{subequations}

Now, the equation resulting from the quadratic terms of \eq{martingaleLinearQuadratic}, i.e. \eq{martingaleQuadratic}, requires
\begin{subequations}\label{eq:twoColoredQuadratic}
\begin{align}
\left( b_{\R}\left(\zeta_{\B}^{-1}-1\right) +a_{\R}\left(\zeta_{\B}-1\right) \right) w_{ji} +\left( b_{\B}\left(\zeta_{\R}^{-1}-1\right) +a_{\B}\left(\zeta_{\R}-1\right) \right) w_{ij} &= 0 ; \quad\left(i\sim\B,\,j\sim\R\right) \\
\left( b_{\B}\left(\zeta_{\R}^{-1}-1\right) +a_{\B}\left(\zeta_{\R}-1\right) \right) w_{ji} +\left( b_{\R}\left(\zeta_{\B}^{-1}-1\right) +a_{\R}\left(\zeta_{\B}-1\right) \right) w_{ij} &= 0 . \quad\left(i\sim\R,\,j\sim\B\right)
\end{align}
\end{subequations}
For the expressions in \eq{zetaBzetaR} to satisfy \eq{twoColoredQuadratic} as well, it must be the case that
\begin{align}
w_{ji}/w_{ij} &= 
\begin{cases}
w_{\B}^{\textrm{in}}/w_{\B}^{\textrm{out}} \ \left( =w_{\R}^{\textrm{out}}/w_{\R}^{\textrm{in}} \right) & i\sim\B ,\ j\sim\R ; \\
w_{\R}^{\textrm{in}}/w_{\R}^{\textrm{out}} \ \left( =w_{\B}^{\textrm{out}}/w_{\B}^{\textrm{in}} \right) & i\sim\R ,\ j\sim\B .
\end{cases} \label{eq:wjiwij}
\end{align}
On a graph satisfying these properties, \eq{fp} says that the fixation probability of a mutant whose initial location is chosen uniformly at random is thus
\begin{align}
\rho_{A} &= \frac{\displaystyle 1-\frac{1}{N}\left(N_{\B}\zeta_{\B}+N_{\R}\zeta_{\R}\right)}{\displaystyle 1-\zeta_{\B}^{N_{\B}}\zeta_{\R}^{N_{\R}}} \nonumber \\
&= \frac{\displaystyle 1-\frac{1}{N}\left\{N_{\B}\frac{b_{\R}w_{\B}^{\textrm{in}}\left(a_{\R}w_{\R}^{\textrm{out}}+b_{\B}w_{\R}^{\textrm{in}}\right)}{a_{\R}w_{\R}^{\textrm{out}}\left(a_{\B}w_{\B}^{\textrm{out}}+b_{\R}w_{\B}^{\textrm{in}}\right)}+N_{\R}\frac{b_{\B}w_{\R}^{\textrm{in}}\left(a_{\B}w_{\B}^{\textrm{out}}+b_{\R}w_{\B}^{\textrm{in}}\right)}{a_{\B}w_{\B}^{\textrm{out}}\left(a_{\R}w_{\R}^{\textrm{out}}+b_{\B}w_{\R}^{\textrm{in}}\right)}\right\}}{\displaystyle 1-\left(\frac{b_{\R}w_{\B}^{\textrm{in}}\left(a_{\R}w_{\R}^{\textrm{out}}+b_{\B}w_{\R}^{\textrm{in}}\right)}{a_{\R}w_{\R}^{\textrm{out}}\left(a_{\B}w_{\B}^{\textrm{out}}+b_{\R}w_{\B}^{\textrm{in}}\right)}\right)^{N_{\B}}\left(\frac{b_{\B}w_{\R}^{\textrm{in}}\left(a_{\B}w_{\B}^{\textrm{out}}+b_{\R}w_{\B}^{\textrm{in}}\right)}{a_{\B}w_{\B}^{\textrm{out}}\left(a_{\R}w_{\R}^{\textrm{out}}+b_{\B}w_{\R}^{\textrm{in}}\right)}\right)^{N_{\R}}} \nonumber \\
&= \frac{\displaystyle 1-\frac{1}{N}\left\{N_{\B}\frac{b_{\R}\left(a_{\R}w_{\B}^{\textrm{in}}+b_{\B}w_{\B}^{\textrm{out}}\right)}{a_{\R}\left(a_{\B}w_{\B}^{\textrm{out}}+b_{\R}w_{\B}^{\textrm{in}}\right)}+N_{\R}\frac{b_{\B}\left(a_{\B}w_{\B}^{\textrm{out}}+b_{\R}w_{\B}^{\textrm{in}}\right)}{a_{\B}\left(a_{\R}w_{\B}^{\textrm{in}}+b_{\B}w_{\B}^{\textrm{out}}\right)}\right\}}{\displaystyle 1-\left(\frac{b_{\R}\left(a_{\R}w_{\B}^{\textrm{in}}+b_{\B}w_{\B}^{\textrm{out}}\right)}{a_{\R}\left(a_{\B}w_{\B}^{\textrm{out}}+b_{\R}w_{\B}^{\textrm{in}}\right)}\right)^{N_{\B}}\left(\frac{b_{\B}\left(a_{\B}w_{\B}^{\textrm{out}}+b_{\R}w_{\B}^{\textrm{in}}\right)}{a_{\B}\left(a_{\R}w_{\B}^{\textrm{in}}+b_{\B}w_{\B}^{\textrm{out}}\right)}\right)^{N_{\R}}} . \label{eq:bithermalFP}
\end{align}

\subsubsection*{Undirected, unweighted graphs}
The process described in the previous section is a modification of the Moran process due to \citet{lieberman2005evolutionary} that allows for general, weighted and directed graphs. When the graph is undirected and unweighted, the Moran process described in the main text can be recovered with an appropriate choice of $\left(w_{ij}\right)_{i,j=1}^{N}$. Suppose that $\left(\Gamma_{ij}\right)_{i,j=1}^{N}$ is the adjacency matrix for such a graph, i.e. $\Gamma_{ij}\in\left\{0,1\right\}$ and $\Gamma_{ij}=\Gamma_{ji}$ for every $i,j=1,\dots ,N$. As before, we assume that there are no self-loops in the graph, so $\Gamma_{ii}=0$ for $i=1,\dots ,N$. We let the probability of transitioning from $i$ to $j$ in one step of a random walk on the graph be $p_{ij}\coloneqq\Gamma_{ij}/\sum_{k=1}^{N}\Gamma_{ik}$. If an individual is selected for reproduction with probability proportional to fitness, and the offspring is subsequently propagated to a random neighboring node, then the probabilities of losing and gaining a mutant at location $i$ in state $\mathbf{x}\in\left\{0,1\right\}^{N}$ are
\begin{subequations}\label{eq:transitionUnDunW}
\begin{align}
P_{i}^{-}\left(\mathbf{x}\right) &= \frac{\sum_{j=1}^{N}x_{i}\left(1-x_{j}\right) b_{j}p_{ji}}{\sum_{j=1}^{N}\left(x_{j}a_{j}+\left(1-x_{j}\right) b_{j}\right)} ; \\
P_{i}^{+}\left(\mathbf{x}\right) &= \frac{\sum_{j=1}^{N}\left(1-x_{i}\right) x_{j}a_{j}p_{ji}}{\sum_{j=1}^{N}\left(x_{j}a_{j}+\left(1-x_{j}\right) b_{j}\right)} ,
\end{align}
\end{subequations}
respectively. Note that these equations are identical to those of \eq{transition} when $\left(w_{ij}\right)_{i,j=1}^{N}=\left(p_{ij}\right)_{i,j=1}^{N}$.

Since $\left(p_{ij}\right)_{i,j=1}^{N}$ is stochastic, the outdegree of every node is $1$. Suppose that this graph is bipartite and properly two-colored, with $N_{\B}$ green nodes and $N_{\R}$ red nodes (see \fig{properColorings}). Following our assumptions in the last section, we assume here as well that all nodes of a given color have the same indegree. Let $p_{\B}^{\textrm{in}}$ and $p_{\R}^{\textrm{in}}$ denote the indegrees of green and red nodes, respectively. By \eq{wjiwij}, we have $p_{ji}=p_{\B}^{\textrm{in}}p_{ij}$ whenever $i\sim\B$ and $j\sim\R$ and $p_{ji}=p_{\R}^{\textrm{in}}p_{ij}$ whenever $i\sim\R$ and $j\sim\B$. Since the graph is connected, these equations imply that there exist $k_{\B}$ and $k_{\R}$ such that $p_{ij}=\Gamma_{ij}/k_{\B}$ when $i\sim\B$ and $p_{ij}=\Gamma_{ij}/k_{\R}$ when $i\sim\R$. In particular, all green nodes have $k_{\B}$ neighbors and all red nodes have $k_{\R}$ neighbors (i.e. the graph is ``biregular").

For this kind of graph, we can simplify the expressions of \eq{zetaBzetaR} to get
\begin{subequations}\label{eq:zetaBzetaR_unweighted}
\begin{align}
\zeta_{\B} &= \frac{b_{\R}\left(a_{\R}k_{\B}+b_{\B}k_{\R}\right)}{a_{\R}\left(a_{\B}k_{\R}+b_{\R}k_{\B}\right)} ; \\
\zeta_{\R} &= \frac{b_{\B}\left(a_{\B}k_{\R}+b_{\R}k_{\B}\right)}{a_{\B}\left(a_{\R}k_{\B}+b_{\B}k_{\R}\right)} .
\end{align}
\end{subequations}
It then follows from \eq{bithermalFP} that the fixation probability of a randomly-placed mutant is
\begin{align}
\rho_{A} &= \frac{\displaystyle 1-\frac{1}{N}\left\{N_{\B}\frac{b_{\R}\left(a_{\R}k_{\B}+b_{\B}k_{\R}\right)}{a_{\R}\left(a_{\B}k_{\R}+b_{\R}k_{\B}\right)}+N_{\R}\frac{b_{\B}\left(a_{\B}k_{\R}+b_{\R}k_{\B}\right)}{a_{\B}\left(a_{\R}k_{\B}+b_{\B}k_{\R}\right)}\right\}}{\displaystyle 1-\left(\frac{b_{\R}\left(a_{\R}k_{\B}+b_{\B}k_{\R}\right)}{a_{\R}\left(a_{\B}k_{\R}+b_{\R}k_{\B}\right)}\right)^{N_{\B}}\left(\frac{b_{\B}\left(a_{\B}k_{\R}+b_{\R}k_{\B}\right)}{a_{\B}\left(a_{\R}k_{\B}+b_{\B}k_{\R}\right)}\right)^{N_{\R}}} . \label{eq:bithermalFP_unweighted}
\end{align}
Swapping the roles of $A$ and $B$ results in a formula for $\rho_{B}$ as well, proving \thm{isothermal}. In particular, any regular graph (meaning $k_{\B}=k_{\R}$, which requires $N_{\B}=N_{\R}$) has the same fixation probabilities as the complete bipartite graph with $N_{\B}$ green nodes and $N_{\R}$ ($=N_{\B}$) red nodes (\fig{properColorings}b).

\begin{example}[star graph]
On the star graph with one red node at the center (see \fig{properColorings}f), we have $N_{\B}=k_{\R}=N-1$ and $N_{\R}=k_{\B}=1$. Plugging these quantities into \eq{bithermalFP_unweighted} gives the formula
\begin{align}
\rho_{A} &= \frac{\displaystyle 1-\frac{1}{N}\left\{\left(N-1\right)\frac{b_{\R}\left(a_{\R}+b_{\B}\left(N-1\right)\right)}{a_{\R}\left(a_{\B}\left(N-1\right)+b_{\R}\right)}+\frac{b_{\B}\left(a_{\B}\left(N-1\right)+b_{\R}\right)}{a_{\B}\left(a_{\R}+b_{\B}\left(N-1\right)\right)}\right\}}{\displaystyle 1-\left(\frac{b_{\R}\left(a_{\R}+b_{\B}\left(N-1\right)\right)}{a_{\R}\left(a_{\B}\left(N-1\right)+b_{\R}\right)}\right)^{N-1}\left(\frac{b_{\B}\left(a_{\B}\left(N-1\right)+b_{\R}\right)}{a_{\B}\left(a_{\R}+b_{\B}\left(N-1\right)\right)}\right)} .
\end{align}
\end{example}

Finally, we conclude with the proof of \thm{selectionCondition}. Suppose that $\rho_{A}=1/N$, the fixation probability of $A$ under neutral drift. By \eq{bithermalFP_unweighted}, with $\zeta_{\B}$ and $\zeta_{\R}$ given by \eq{zetaBzetaR_unweighted}, we have
\begin{align}
N_{\B}\left(1-\zeta_{\B}\right) + N_{\R}\left(1-\zeta_{\R}\right) &= 1 - \zeta_{\B}^{N_{\B}}\zeta_{\R}^{N_{\R}} . \label{eq:neutralSI}
\end{align}
Consider the function $f\left(x,y\right)\coloneqq N_{\B}\left(1-x\right) + N_{\R}\left(1-y\right) - 1 + x^{N_{\B}}y^{N_{\R}}$, which satisfies $f\left(1,1\right) =0$. If $\left(\zeta_{\B},\zeta_{\R}\right)\neq\left(1,1\right)$ and $f\left(\zeta_{\B},\zeta_{\R}\right) =0$, then the function $g\left(t\right)\coloneqq f\left(1-t+t\zeta_{\B},1-t+t\zeta_{\R}\right)$ is differentiable and vanishes at both $t=0$ and $t=1$. By Rolle's theorem, there must exist $t^{\ast}\in\left(0,1\right)$ for which $g'\left(t^{\ast}\right) =0$. Letting $x^{\ast}\coloneqq 1-t^{\ast}+t^{\ast}\zeta_{\B}$ and $y^{\ast}\coloneqq 1-t^{\ast}+t^{\ast}\zeta_{\R}$ be the corresponding values of $x$ and $y$, we see that
\begin{align}
g'\left(t^{\ast}\right) &= \frac{\partial f}{\partial x}\left(x^{\ast},y^{\ast}\right)\left(\zeta_{\B}-1\right) + \frac{\partial f}{\partial y}\left(x^{\ast},y^{\ast}\right)\left(\zeta_{\R}-1\right) \nonumber \\
&= -N_{\B}\left(1-\left(x^{\ast}\right)^{N_{\B}-1}\left(y^{\ast}\right)^{N_{\R}}\right)\left(\zeta_{\B}-1\right) - N_{\R}\left(1-\left(x^{\ast}\right)^{N_{\B}}\left(y^{\ast}\right)^{N_{\R}-1}\right)\left(\zeta_{\R}-1\right) .
\end{align}
Since the numerators and denominators in \eq{zetaBzetaR_unweighted} are positive, and since we have the identities
\begin{subequations}
\begin{align}
a_{\R}\left(a_{\B}k_{\R}+b_{\R}k_{\B}\right) - b_{\R}\left(a_{\R}k_{\B}+b_{\B}k_{\R}\right) &= k_{\R}\left(a_{\B}a_{\R}-b_{\B}b_{\R}\right) ; \\
a_{\B}\left(a_{\R}k_{\B}+b_{\B}k_{\R}\right) - b_{\B}\left(a_{\B}k_{\R}+b_{\R}k_{\B}\right) &= k_{\B}\left(a_{\B}a_{\R}-b_{\B}b_{\R}\right) ,
\end{align}
\end{subequations}
it follows that $1-\zeta_{\B}$ and $1-\zeta_{\R}$ must have the same sign (positive, negative, or zero). Since $\left(1-x^{\ast},1-y^{\ast}\right)$ lies in the same quadrant as $\left(1-\zeta_{\B},1-\zeta_{\R}\right)$, it cannot be true that $g'\left(t^{\ast}\right) =0$. Therefore, \eq{neutralSI} is satisfied if and only if $\zeta_{\B}=\zeta_{\R}=1$, which in turn happens if and only if $a_{\B}a_{\R}=b_{\B}b_{\R}$. For bipartite, properly two-colored, biregular graphs, we see that $\rho_{A}=1/N$ if and only if $a_{\B}a_{\R}=b_{\B}b_{\R}$. By similar reasoning, we find that $\rho_{A}>1/N>\rho_{B}$ when $a_{\B}a_{\R}>b_{\B}b_{\R}$ and $\rho_{A}<1/N<\rho_{B}$ when $a_{\B}a_{\R}<b_{\B}b_{\R}$.

\section*{Acknowledgments}
We thank Igor Erovenko and two anonymous referees for many helpful comments on an earlier version of this paper. The authors gratefully acknowledge support from the Army Research Laboratory (grant W911NF-18-2-0265), the Bill \& Melinda Gates Foundation (grant OPP1148627), and the NVIDIA Corporation.

\end{document}